\documentstyle{article}
\begin{document}
 
\title{Nomenclature of The Galactic Center Radio Sources} \author
{Patrick Palmer {\it (University of Chicago)} \and W. Miller Goss {\it
(National Radio Astronomy Observatory)$^1$}}

\date{in: Galactic Center Newsletter - GCNEWS (1996), Vol. 2, p. 2\\
(http://www.astro.umd.edu/$\sim$gcnews/gcnews/Vol.2/article.html)}
\maketitle

\begin{abstract}
Since its identification as the Galactic Center, the radio source
Sagittarius A has been a target of intense research. Due to the high
density of sources in the Galactic Center and differing observing
techniques, the nomenclature of sources in this region has changed
over the years, with sources having several names, as well as the same
names being used for different sources. We review this historical
evolution in the context of current and previous scientific
discussions, and outline how, why and when some of the commonly accepted
names of Galactic Center sources were established.
\end{abstract}

The discovery of the dominant source in the Galactic center, Sagittarius
A, in the era 1951 -- 1960 has been described by Goss and McGee at the recent
ESO/CTIO conference$^2$. Due to the high density of sources in the Galactic
center and the wide range of spectral indices, the recognition of the many
components of Sgr A was dependent on both resolution and observing
frequency.

In some sense Karl Jansky and Grote Reber did discover the Galactic center
radio source in the 1930's -- 1940's. However, with their poor resolutions
(beams of many degrees) the radio emission that was detected was the Galactic
background peak near the Galactic center.  The recognition that 
discrete radio sources existed (other than the Sun) emerged in the era 1946 --
1948$^3$; until that era radio emission was regarded as coming from a
general background$^4$.

The realization that Sgr A is a discrete radio source at the Galactic
center was first made by Jack Piddington and Harry Minnett$^5$ .  These authors
had the handicap that they published in a rather obscure journal --
Australia was even more isolated from Europe and the US in the 1950's than
at present. Often the credit for the association of Sgr A with the Galactic
nucleus is given to McGee and Bolton based on their 1954 paper in {\it
Nature}$^6$.  Of course, the fact that {\it Nature} was more widely read
than the Australian publication did help.  However, many famous astronomers
of the 20th century played a role in the preparation, refereeing, and
publication of the 1954 paper: Baade, Oort, van de Hulst, Pawsey, Mills,
Kerr, Bracewell and Shain.  Goss and McGee have corrected the previously
published record on the details of the publication of the 1954 {\it Nature}
paper. Bernie Mills has been especially helpful in reconstructing the
atmosphere at the CSIRO Division of Radiophysics in Sydney, NSW, Australia in
the early 1950's and the understanding of the nature of radio sources at
that time.  In the late 1950's the association of Sgr A with the center of
the Milky Way was generally accepted$^7$.

Dick McGee pointed out to us that in those early years the name Sgr A was
{\it not} used in Australia . This is rather puzzling since John Bolton and
his collaborators (Bruce Slee, Gordon Stanley and Kevin Westfold) had given
the names Taurus A, Centarus A, and Virgo A to other early radio sources
discovered at Dover Heights in Sydney. Goss has tried to find the earliest
reference to Sgr A (can anyone find an earlier reference ?).  The earliest
found is in {\it Sky and Telescope} in 1954$^8$ in a summary of a paper
given at the June 1954 American Astronomical Society Meeting in Ann Arbor,
Michigan by John Kraus and H. C. Ko.  John Kraus has recently written: 
``... whether or not I actually \lq invented\rq\ the name Sagittarius A , I was
certainly one of the first to use it consistently. Then more recently I
took the naming a step further by calling it J1 for Jansky One. See page
313 of my book {\it Big Ear Two} ....'' $^9$ As we know J1 did not catch on!

The names of the discrete sources around Sgr A have a complex and confusing
history.  One might ask why bother with it: the names based on galactic
coordinates are perfectly sufficient and unambiguous.  First, if one wishes
to read the older literature, one must be aware of certain pitfalls; but,
perhaps the more important result of this study is that it reveals the
importance of private communications and unpublished manuscripts in the
development of this field.

The first map of the Galactic center region with adequate resolution to
resolve several distinct sources around Sgr A was made by Frank Drake in
1959 with the 85 foot telescope at Green Bank at 3.75 cm$^{10}$. Drake
himself never named the components$^{11}$.  However, he distributed the map
widely, including sending a copy to James Lequeux for use in his thesis.
Lequeux recalls$^{12}$ adding the designations A for Sgr A, and B1, B2, and
B3$^{13}$ for the other three prominent discrete sources (ordered by their
distance from Sgr A) to Drake's figure for use in his 1962 paper in {\it
Annales d'Astrophysique}$^{14}$ . In modern terms, sources B1, B2, and B3
correspond to G0.2-0.0, the blend of G0.5-0.0 and G0.7-0.0, and
G359.4-0.1. (Kraus and Ko had already defined a Sgr B in 1954$^8$; however,
this source at $\ell\sim4\deg$ \ does not appear elsewhere in the
literature.)

Soviet astronomers, especially Yuri Parijskij, also were studying the
Galactic center at this time$^{15}$, and they also proposed names for the
components.  However, because their source names were not picked up by
others, no attempt will be made here to trace them.  Drake notes
facetiously that it is a pity that these names were not used because the
sources on opposite sides of Sgr A (B2 and B3) were interpreted as part of
a ring which was named the ``Drake Ring''$^{11}$.

Relatively little use was made of Lequeux's names.  For example, in Cooper
and Price's 10 cm map made at Parkes in 1962, no names were attached to the
sources$^{16}$. Rougoor, however, used Lequeux's names in his 1964 paper on
the nuclear region of the Galaxy$^{17}$.  

A surprisingly influential unpublished paper helped disseminate the Lequeux
names in Jaunary 1965.  Dennis Downes, then an undergraduate at Harvard,
wrote a term paper on Galactic center radio emission for a course given by
Alan Maxwell$^{18}$.  In this paper he used Lequeux's naming system and
added B4 for a source near $\ell=1.1\deg$ \ that was on Cooper and Price's
map but beyond the edge of Drake's map.  This paper was distributed to
other radio astronomers at Harvard; and, because it was found to be very
useful by them, 50 - 100 copies were made and distributed to other radio
astronomers around the world$^{19}$.  However, when this paper was
published in 1966 in collaboration with Alan Maxwell$^{20}$, the Sgr Bn
form of names was discarded and names based on Galactic coordinates were
used.

The shift to names based on Galactic coordinates had been discussed for
some time, but it was most clearly proposed by Mezger and Henderson in
1967$^{21}$.  In fact, the Galactic center region was one of their prime
examples that something else should be done: ``... the example of the
Sagittarius region, where designations A, B, C, etc., are used
concurrently with designations A, B1, B2, etc., shows clearly that this
type of nomenclature will lead to serious confusion as more results of high
resolution studies ... become available.''

The last appearance of B1 in the literature for more than 20 years (and the
only appearance of B4) is found in a paper on OH by Palmer and Zuckerman in
1967$^{22}$.  The name B2 persisted (although it silently migrated from
$\ell=0.6\deg$ \ -- the centroid of the low resolution blend -- to
$\ell=0.7\deg$ ), probably because of the almost continuous interest in
this source due to the various molecular discoveries.  Otherwise the system
of naming components by their Galactic coordinates triumphed among the
radio astronomers.

A new attempt to name Galactic center sources was made by Hoffmann,
Fredrick, and Emery in 1971$^{23}$ for their 100 micron survey of this
region.  They named sources A, B2, C, D, and E.  Sgr A and B2
corresponded to the radio astronomer's sources with the same names, but
they introduced C (corresponding to Lequeux's B3 at $\ell=359.4\deg$), D
(corresponding to Downes' B4 at $\ell=1.1\deg$), and E (only later studied by
radio astronomers at $\ell=358.4\deg$).  Sgr B1 was lost.

Sgr B1 reappeared in 1986 in a proposal for recombination line studies at
the VLA by Palmer, Yusef-Zadeh, Goss, Lasenby, \& Lasenby$^{24}$.  The name
Sgr B1 was apparently suggested by Yusef-Zadeh.  He knew that a source name
Sgr B1 had been used in the past, but had been unable to find any definite
information about it, and made the apparently straightforward
identification of G0.5-0.0 with Sgr B1$^{25}$.  The observations were carried
out, as were other observations of the region, and the name Sgr B1 came
back into use$^{26}$.

However, Sgr B1 was not the source so named by Lequeux!  In modern terms,
Lequeux's Sgr B1 is the source now called G0.2-0.0.  The source called Sgr
B2 in the early papers was a blend of G0.5-0.0 and G0.7-0.0, the sources
now called Sgr B1 and Sgr B2.  (Palmer should have known better, but did
not think about the contradiction between the OH spectra for Sgr B1 and Sgr B2
in the paper referenced in note 22 if Sgr B1 had been at the current
position until discussion leading to preparation of this report.)

What of the future for names of Galactic sources?  It is clear that there
is no simple solution.  The G -- type names are indeed unambiguous; but at
the time they were proposed, the extent to which Galactic sources would
continue to resolve was not foreseen.  For example, Sgr B2 is now known to
contain almost 60 components$^{27}$.  In order to manage this complexity,
one needs some larger organizing principles which the letter -- type names
provide.  It seems that for practical reasons both types of names will
continue to coexist.

We are indebted to very helpful correspondence from a number of
individuals, some of whom were ``bugged'' many times for details.  We
especially wish to thank Dennis Downes, Frank Drake, James Lequeux, Harvey
Liszt, John Kraus, Dick McGee, David Mehringer, Peter Mezger, Mark Morris,
and Farhad Yusef-Zadeh.

\vskip10pt
\centerline{Notes}

\begin{itemize}
\item[1] The National Radio Astronomy Observatory is a
facility of the National Science Foundation operated under 
cooperative agreement by Associated Universities, Inc.
\item[2] Goss, W. M. \& McGee, R. X. 1996, in {\it Proceedings of the 4th
ESO/CTIO Workshop: The Galactic Center} ed Gredel and Schommer, in press.
\item[3] Beginning with  Hey,  Parsons \&  Phillips,  (Nature,
158, 234 (1946)) who write ``It appears probable that such variations could
only originate from a small number of discrete sources.''; followed by
Bolton and Stanley (Australian J. Sci. Res., Ser A, 1, 58 (1948)) and Ryle
and Smith (Nature, 162, 462 (1948)) who provided the earliest source
diameter measurements.
\item[4] For example, Pawsey, Payne-Scott \& McCready (Nature, 157, 158
(1946)) argued that the observed radio emission was the sum of emission
from large numbers of stars emitting radio waves by the same (unknown)
mechanism as sunspots; while Greenstein, Henyey, \& Kennan (Nature, 157, 805
(1946)) defended their view that it was free-free emission from the
interstellar medium.
\item[5] Piddington, J. H. \& Minnett, H. C. 1951, Australian J.  
Sci. Res, Ser A, 4, 495.
\item[6] McGee, R. X. \& Bolton, J. G. 1954, Nature, 173, 985.
\item[7] Representative examples of this acceptance are: the title of an 
article ``The Radio Position of the Galactic Nucleus'' (Kraus, J. D. \& Ko,
H. C 1955, Ap. J., 122, 139); a note on the source W24 ``believed to be the
Galactic nucleus'' (Westerhout, G. 1958, B. A. N., 14, 215), and finally
the five papers published in MNRAS giving reasons for the redefinition 
of the Galactic coordinate system and the actual form which the new
definition takes.  These papers begin with  Blaauw, Gum, Pawsey, and
Westerhout (MNRAS, 121, 123 (1960)) which states: ``We shall, on the basis of
evidence presented in Paper V, assume that Sagittarius A is located at the
Galactic centre.''
\item[8] (uncredited news story) 1954, Sky \& Telescope, 14, 22; subsequently
published in: Kraus, J. D. \& Ko, H. C. 1955, Ap. J., 175, 159.
\item[9] J. Kraus, letter to W. M. Goss, May 7, 1996.

\item[{10}] This map was never published in the refereed literature, but did
appear in the NRAO Annual Report for 1959.
\item[{11}] F. Drake, telephone conversation, July 10, 1996.
\item[{12}] J. Lequeux, letter to W. M. Goss, June 13, 1996.
\item[{13}] Lequeux used the numbers as subscripts.  However, almost from the
 beginning, there was no consistency about whether or not the numbers were
 subscripts; consequently, there is little point trying to decide on a
 ``correct'' way to write them.
\item[{14}] Lequeux, J. 1962, Annales d'Astrophysique, 25, 221.
\item[{15}] See for example, Parijskij, Y. 1959, Soviet Physics -- Doklady, 4, 1172.
\item[{16}] This map was first shown by Kerr (Sky \& Telescope, 24, 254 (1962)).  It
was later published by  Cooper, B. F. C. \& Price, R. M. 1964 in {\it The
Galaxy and the Magellanic Clouds}, ed F. J. Kerr \& A. W. Rodgers
(Aust. Acad. Sci.), p. 1964.
\item[{17}] Rougoor, G. W. 1964, B. A. N., 17, 381.

\item[{18}] D. Downes, unpublished manuscript, January, 1965; revised August,
1965.  
\item[{19}] D. Downes, letter to W. M. Goss, July 10, 1996.  A copy was sent to
 J. G. Bolton who acknowledged receipt of the paper in a letter to Downes on
 October 15, 1965.
\item[{20}] Downes, D. \& Maxwell, A. 1966, Ap. J., 146, 653.
\item[{21}] Mezger, P. G. \& Henderson, A. P. 1967, Ap. J., 147, 471.
\item[{22}] Palmer, P. \& Zuckerman, B. 1967, Ap. J., 148, 727.
\item[{23}] Hoffmann, W. F., Fredrick, C. L. \& Emery, R. J. 1971, Ap. J., 164,
L23.

\item[{24}] Proposal AP139, received Dec, 19, 1986; VLA proposal archive.
\item[{25}] F. Yusef-Zadeh, telephone conversation, July 10, 1996.
\item[{26}] See, for example, Liszt, H. L. 1988, in {\it Galactic and Extragalactic
Radio Astronomy} ed G. L. Verschuur \& K. I. Kellermann (Springer: New
York), p. 359; Morris, M. 1989, in {\it The Center of The Galaxy} ed
M. Morris (Kluwer: Dordrecht), p. 171; and Mehringer, D. M., Yusef-Zadeh,
F., Palmer, P. \& Goss, W. M. 1992, Ap. J., 401, 168.
\item[{27}] Gaume, R. A., Claussen, M. J., De Pree, C. G., Goss, W. M., \&
Mehringer, D. M. 1995, Ap. J., 449, 663.
\item[{28}] The interest of W. M. Goss in this topic began during visits with
John \& Letty Bolton in Buderim, Queensland, Australia in September, 1988
and November, 1992.  The last visit was a short time before John Bolton's
death on July 6, 1993.
\end{itemize}
\end{document}